\documentstyle[aps,epsf,prc]{revtex}

\begin{document}
\title{Scalar meson production in nucleon-nucleon collisions near threshold}
\author{Michail P. Rekalo \footnote{ Permanent address:
\it National Science Center KFTI, 310108 Kharkov, Ukraine}
}
\address{Middle East Technical University, 
Physics Department, Ankara 06531, Turkey}
\author{Egle Tomasi-Gustafsson}
\address{\it DAPNIA/SPhN, CEA/Saclay, 91191 Gif-sur-Yvette Cedex, 
France}
\date{\today}

\maketitle
\begin{abstract}
We establish the model-independent spin structure of the matrix elements for the near-threshold scalar meson production in $pp-$ and $np$-collisions, when the final particles are emitted in S-state. Polarization phenomena are derived in a general form. The properties of the $t-$channel dynamics, which is based on different meson exchanges, are studied in terms of the $s-$channel parametrization of the matrix element. The prediction of a 'realistic' model, based on $\pi+\sigma$-exchanges are also presented. 
\end{abstract}
\section{Introduction}
The strong and electromagnetic decays of the scalar mesons $S$ ($S= \sigma$, $f_0$, and $a_0$) are recently object of large theoretical and experimental interest. The structure of these mesons is not yet fully understood and the coupling constants of their decays not determined (see \cite{Cl93,Ge94,Ja95,An95,Ma00,Na00,Cl00,Mo02} and refs. herein). As an example, the electromagnetic constants $g_{\rho\sigma\gamma}$ and $g_{\omega\sigma\gamma}$ are very important for the solution of different problems in in hadron electrodynamics. In the near threshold region, the constant $g_{\rho\sigma\gamma}$ drives the $\sigma$-contribution to the differential cross section of the process $\gamma+p\to p+p+\rho^0$ \cite{Fr00,Zh98,Ti99}. The coupling constant  
$g_{\omega\sigma\gamma}$ is important for the estimation of the effects of the meson exchange currents 
for the deuteron electromagnetic form factors \cite{Ch74,Hu90,VO95}, particularly in the region of large momentum transfer. Both these constants play a major role in the interpretation \cite{Mi00} of the HERMES effect \cite{Ac99}, concerning the anomalous behavior of the electroproduction cross section on nuclei at low 
$Q^2$ - where the cross section is enhanced for longitudinally polarized virtual photons and depleted for transversally polarized photons. Moreover, the constants $g_{\rho\sigma\gamma}$ and $g_{\omega\sigma\gamma}$ enter in the interpretation of different radiative decays of vector mesons, like $\rho^0(\omega)\to \pi^0\pi^0\gamma$ and $\rho^0(\omega)\to \pi^+\pi^-\gamma$ 
\cite{Ac00}. 

Two-photon decays of scalar mesons, $S\to 2\gamma$, which  are important for the estimation of the corresponding $t-$channel contributions to the amplitudes of real and virtual Compton scattering on nucleons \cite{An99,Bo98,Li90,Va96}, could be estimated on the basis of the $g_{V\sigma\gamma}$ coupling constants, in the VDM approach.

In the framework of the effective Lagrangian approach, the calculation of the cross section for scalar meson photo- and electro-production on nucleons requires the knowledge of the  $g_{V\sigma\gamma}$ and $g_{NNS}$-coupling constants. The strong coupling constants   $g_{NNS}$ \cite{El87,Ma95} enter in various calculations in hadron dynamics, not only with respect to the NN-potential, but also for different observables for processes like $\pi+N\to S+N$ \cite{Br99,Gr00}, $N+N\to S+d$ \cite{Gr00,Mu01,Ku00,Ku02}, and $N+N\to S+N+N$ \cite{Br99}. 

The feasibility of the experimental study of scalar meson production depends essentially on the nature and on the rate of the decay. If, for example, the decay $f_0\to K\overline{K}$ or $a_0\to K\overline{K}$ dominate, these mesons could be observed in the $K\overline{K}$ effective mass distribution, close to the kaon mass, as a resonant contribution. Such effects have been observed in $\pi N$-collisions \cite{An95,Da69,Be67,Pa75,Ca78,Co80,Et82} . The interpretation of the data needs an adequate theoretical approach for the process $\pi +N\to N+S^0$. It was shown that the reaction $p+p\to p+p+K^+K^-$ in the kinematical conditions of the DISTO \cite{Disto} or at  COSY the presence of the $f_0$ signal is hindered by a large background. Therefore, in principle, the process $n+p\to n+p+K^+K^-$  could be more favorable, because the cross section is one order higher, whereas the background is comparable with respect to $pp$-collisions.

Note that the experimental study of $K^+K^-$-production in the process $p+p\to p+p+K^++K^-$, at an energy excess Q=17 MeV over threshold \cite{Qu01}, gives the following value: $\sigma(pp\to pp f_0)=\left (1.84\pm 0.29^{+0.25}_{-0.35}\right) $ nb, including the statistical and systematic errors. The extension of this study is foreseen \cite{Oe88,Bu01,Mo01}.

In this paper we derive the most general and model independent properties for the processes of scalar meson production in $NN-$collisions, in the threshold region, where the theoretical analysis is essentially simplified. The spin structure of the corresponding matrix elements contains a set of sixteen independent amplitudes in the general case, eight amplitudes for coplanar kinematics and only three independent amplitudes in the threshold region.

This paper is organized as follows. In Section II, using the selection rules with respect to Pauli principle, P-parity and total angular momentum, we establish the spin structure of the matrix elements for the processes $p+p\to p+p+S^0$ and $n+p\to n+p+S^0$ and analyze the polarization phenomena for these processes in a model independent way. In Section III we transform (using the two-component Fierz transformation) different $t-$channel contributions, which are described by a definite set of Feynman diagrams, to the universal $s$-channel parametrization and find the expression for the corresponding partial threshold amplitudes. In Section IV we discuss the predictions of a realistic model based on $\sigma+\pi$-exchanges. A short discussion of final state interactions (FSI) is done in Section V. In the Conclusions we summarize the results obtained.

\section{Spin structure of the threshold matrix element and polarization observables }
The spin structure of the matrix elements for scalar meson production in $NN-$collisions, $N+N\to N+N+S^0$, in the threshold region is determined by the selection rules 
with respect to  $P$-parity, total angular momentum and by the Pauli principle. The threshold region, where all final particles in $N+N\to N+N+S^0$ are produced in relative S-state, can be rigorously described by a formalism based on the two-component nucleon spinor parametrization of the corresponding matrix element. This formalism can be built in model independent way \cite{ETG02}, and has been previously successfully applied to vector \cite{Re1}, pseudoscalar \cite{Re2}, and strange \cite{Re3} particle production in nucleon-nucleon collisions.

Let us explicitly derive the matrix element for the processes $p+p\to p+p+S^0$, where $S^0$ denotes a neutral scalar meson, $S^0=\sigma$, $f_0$ or $a_0^0$. Due to the Pauli principle, at the reaction threshold, only quantum numbers $j^P=0^+$ are allowed,  where $j$ is the total angular momentum  and  $P$ is the $P$-parity of the colliding protons. Only a single partial transition can take place, corresponding to:
\begin{equation}
S_i=0,~\ell=0~\to ~j^{ P}=0^+~\to S_f=0, 
\label{eq:eq1}
\end{equation}
where $S_i$ ($S_f$)  is the total spin of the initial (final) protons and 
$\ell$ is the angular orbital momentum of the colliding protons.

The spin structure of the threshold matrix element for the transition (\ref{eq:eq1}), in the CMS of the considered reaction,  can be parametrized in the following general form:
\begin{equation}
{\cal M}_{pp}=g(\chi^{\dagger}_4\sigma_y \tilde{\chi}^{\dagger}_3 )
(\tilde\chi_2\sigma_y {\chi}_1 ),
\label{eq:mpp}
\end{equation}
where $\chi_1$ and $\chi_2$ ($\chi_3$ and $\chi_4$) are the
two-component spinors of the initial (final) protons; $g$ is the threshold partial amplitude, describing the singlet-singlet transition in the $pp$-system (with scalar meson production in the $S$-state), which is generally a complex function of three independent energies:
$W$ (the total invariant energy of the colliding particles), $E_1$ and  $E_2$ (the energies of the scattered protons). So, all the dynamics of the considered process is included in the amplitude $g$, but the exact form (\ref{eq:mpp}) of the spin structure of the matrix element results from a generalized quantum mechanical kinematics.

The Pauli matrix $\sigma_y$, in the parametrization (\ref{eq:mpp}), insures the correct transformation properties of the corresponding two-component spinor products, relative to rotation. 

The presence of a single amplitude in (\ref{eq:mpp}) implies that the spin directions of all the  protons are fixed, with  definite relative angles and fixed modules. In other words, all polarization phenomena for $p+p\to p+p+S^0$ at threshold can be predicted without knowing the amplitude $g$, i.e. {\it in model independent way}. Moreover, these polarization effects are the same  for any scalar meson: $\sigma$, $f_0$ or $a_0^0$. However, the absolute value of the cross section, which depends on $|g^2|$, is different for different scalar mesons.

All T-odd polarization observables (i.e. one-spin and three-spin polarization correlations) are identically zero. The dependence of the cross section on the polarizations $\vec P_1$ and $\vec P_2$ of the colliding protons can be written as:
$$
\displaystyle\frac{d\sigma}{d\omega}(\vec P_1,\vec P_2) =\left (\displaystyle\frac{d\sigma}{d\omega}\right )_0\left (1-\vec P_1\cdot \vec P_2\right ),
$$
(typical for the singlet $pp-$interaction), where $d\omega$ is the phase space volume element for the three-particles final state and  $({d\sigma}/{d\omega})_0$ is the differential cross section with all unpolarized protons in initial and final states. 

All polarization transfer coefficients, characterizing the dependence of the polarization of any final proton from the polarization of the initial proton, vanish also.

The situation is very different in case of $np-$ collisions, $n+p\to n+p+S^0$, in the threshold region, where there are three allowed transitions:
\begin{eqnarray*}
S_i=0,~\ell=0~&\to& ~j^{ P}=0^+\to S_f=0,  \\
S_i=1,~\ell=0~&\to& ~j^{ P}=1^+\to S_f=1,\\
S_i=1,~\ell=2~&\to& ~j^{ P}=1^+\to S_f=1. 
\end{eqnarray*}
The corresponding matrix element can be written in the following form:
\begin{eqnarray}
& {\cal M}_{np}=&g_1({\chi}^{\dagger}_4\sigma_y \tilde\chi^{\dagger}_3)
(\tilde\chi_2\sigma_y \chi_1)\nonumber \\
&&+g_2\left [\chi^{\dagger}_4(\sigma_a-\hat{k_a}
\vec\sigma\cdot\hat{\vec k })\sigma_y\tilde\chi^{\dagger}_3\right ]
\left [(\tilde\chi_2\sigma_y(\sigma_a-\hat{k_a}
\vec\sigma\cdot\hat{\vec k })\chi_1\right ]
\label{eq:mel}\\
&&+g_3(\chi^{\dagger}_4\vec\sigma\cdot\hat{\vec k }\sigma_y\tilde\chi^{\dagger}_3)(
\tilde\chi_2\sigma_y\vec\sigma\cdot\hat{\vec k }\chi_1),
\nonumber
\end{eqnarray}
where $\hat{\vec k }$ is the unit vector along the three-momentum of the initial neutron beam, $\vec\sigma=(\sigma_x,\sigma_y,\sigma_z)$ is the standard set of Pauli matrices, and $g_1-g_3$ are the partial amplitudes for the $np$-interaction. 
Due to the isotopic invariance of the strong interaction the following relation holds:
\begin{equation}
g_1=\displaystyle\frac{1}{2}g,
\label{eq:eq4}
\end{equation}
i.e. the triplet-triplet amplitudes $g_2$ and $g_3$ are present only in $np$-collisions. Therefore polarization phenomena in $n+p\to n+p+S^0$ collisions are more complicated then in $pp$-collisions. There is no universality here, polarization phenomena are different for the different scalar mesons. There is, however one general feature: all one-spin observables vanish, for $n+p\to n+p+S^0$, in particular the T-odd polarization of the final nucleons, emitted in the collisions of unpolarized nucleons and the analyzing powers in $\vec n+p$- or $n+\vec p$-collisions.

The dependence of the differential cross section for $\vec n+\vec p$-collisions can be written in the following general form, which is correct near threshold:
\begin{equation}
\displaystyle\frac{d\sigma}{d\omega}(\vec P_1,\vec P_2) =
\left (\displaystyle\frac{d\sigma}{d\omega}\right )_0\left (1+{\cal A}_1\vec P_1\cdot \vec P_2+{\cal A}_2\hat{\vec k }\cdot \vec P_1\hat{\vec k }\cdot \vec P_2\right ),
\label{eq:sig}
\end{equation}
where the real spin correlation coefficients ${\cal A}_1$ and ${\cal A}_2$ are determined by the following formulas (in terms of the partial threshold amplitudes $g_i$, $i=1-3$):
\begin{eqnarray}
&{\cal A}_1\left (\displaystyle\frac{d\sigma}{d\omega}
\right )_0=&-|g_1|^2+|g_3|^2,\nonumber\\
&{\cal A}_2\left (\displaystyle\frac{d\sigma}{d\omega}
\right )_0=&2(|g_2|^2-|g_3|^2).
\label{eq:eq6}
\end{eqnarray}
The amplitudes $g_i$ are normalized in such way that:
\begin{equation}
\left (\displaystyle\frac{d\sigma}{d\omega}\right )_0=|g_1|^2+2|g_2|^2+|g_3|^2,
\label{eq:eq7}
\end{equation}
where $({d\sigma}/{d\omega})_0$ is the differential cross section with unpolarized particles.
So, from Eqs. (\ref{eq:eq6}) and (\ref{eq:eq7}) one can find:
\begin{eqnarray}
4|g_1|^2&=&(1-3{\cal A}_1-{\cal A}_2)\left (\displaystyle\frac{d\sigma}{d\omega}\right )_0,\nonumber\\
4|g_2|^2&=&(1+{\cal A}_1+{\cal A}_2)\left (\displaystyle\frac{d\sigma}{d\omega}\right )_0,
\label{eq:eq8}\\
4|g_3|^2&=&(1+{\cal A}_1-{\cal A}_2)\left (\displaystyle\frac{d\sigma}{d\omega}\right )_0,\nonumber
\end{eqnarray}
which shows that the moduli of the three threshold amplitudes for $n+p\to n+p+S^0$ can be determined by the measurement of 
${\cal A}_1$, ${\cal A}_2$, and $\left ({d\sigma}/{d\omega}\right )_0$.  Taking into account the isotopic relation (\ref{eq:eq4}), one obtains:
$${\cal R}=\displaystyle\frac{\left ({d\sigma}_{np}/{d\omega}\right )_0}{\left ({d\sigma}_{pp}/{d\omega}\right )_0}=1-{\cal A}_1-{\cal A}_2.$$

The dependence of the polarization $\vec P_3$ of the final neutron on the polarization  $\vec P_1$ of the initial neutron can be parametrized as:
\begin{equation}
\vec P_3=p_1\vec P_1+p_2\hat{\vec k }(\hat{\vec k }\cdot \vec P_1),
\label{eq:eq9}
\end{equation}
where $p_1$ and $p_2$ are the spin transfer coefficients, which can be expressed in terms of the partial amplitudes $g_i$:
\begin{equation}
p_1=-\displaystyle\frac{2{\cal R}e g_2(g_1-g_3)^*}{|g_1|^2+2|g_2|^2+|g_3|^2},
~p_2=\displaystyle\frac{2|g_2|^2-2{\cal R}e\left [g_1g_3^*-g_2(g_1-g_3)^*\right ]}{|g_1|^2+2|g_2|^2+|g_3|^2}.
\label{eq:eq10}
\end{equation}
The dependence of the polarization $\vec P_4$ of the final proton on the polarization  $\vec P_1$ of the initial neutron can be also be parametrized in terms of two real coefficients:
\begin{equation}
\vec P_4=p_3\vec P_1+p_4\hat{\vec k }(\hat{\vec k }\cdot \vec P_1)
\label{eq:eq11}
\end{equation}
with the following formulas for $p_3$ and $p_4$:
\begin{equation}
p_3=\displaystyle\frac{2{\cal R}eg_2( g_1+g_3)^*}{|g_1|^2+2|g_2|^2+|g_3|^2},
~p_4=\displaystyle\frac{2|g_2|^2+2{\cal R}e\left [g_1g_3^*-g_2(g_1+g_3)^*\right ]}{|g_1|^2+2|g_2|^2+|g_3|^2}.
\label{eq:eq12}
\end{equation}
Comparing Eqs. (\ref{eq:eq6}), (\ref{eq:eq7}), (\ref{eq:eq10}), and (\ref{eq:eq12}), one can find the following relation between these polarization observables:
\begin{equation}
p_1+p_2+p_3+p_4=1+{\cal A}_1+{\cal A}_2, 
\label{eq:eq13}
\end{equation}
which has to be verified by any model describing this process.

Note that the isovector $a_0$-meson can be produced in the following processes of the $NN$-interaction:
\begin{eqnarray*}
&p+p&\to p+p+a_0^0,   \\
&p+p&\to n+p+a_0^+,\\
&n+p&\to p+p+a_0^-,\\
&n+p&\to n+p+a_0^0.
\end{eqnarray*}
It is possible to show that the production of the charged $a_0^\pm$-mesons -in the near threshold regime is characterized by a single amplitude, with the spin structure as in Eq. (\ref{eq:mpp}), where the following relations hold:
$$g(pp\to npa_0^+)=-g(np\to pp a_0^-)=\displaystyle\frac{1}{\sqrt{2}} g (np\to pp a_0^0).$$
Therefore the previous statements on polarization phenomena for  $p+p\to p+p+S^0$ apply also to the production of the charged $a_0^\pm$-mesons.

\section{The dynamics for the $\lowercase{t}$-channel}

The standard dynamics for different processes of meson production in NN-collisions, $N+N\to N+N+V$, $N+N\to N+N+P$, and $N+N\to N+N+S^0$, (where $P$, $S$, and $V$ denote pseudoscalar, scalar and vector mesons, respectively) is based on the consideration of mesonic exchanges in $t-$channel, such as $\pi$, $\eta$, $\sigma$, $a_0$, $\rho$, $\omega$, etc. In such approach one has to know the meson-nucleon coupling constants, $PNN$, $SNN$, and $VNN$ and the amplitudes of different subprocesses, such as :
\begin{eqnarray}
&P^*+N&\to N+S, \nonumber  \\
&S^*+N&\to N+S,\label{eq:eq14} \\
&V^*+N&\to N+S, \nonumber  
\end{eqnarray}
where the index '$^*$' denotes virtual mesons - with space-like four-momenta. In principle some information exist on the coupling constants, but the threshold amplitudes  for the processes (\ref{eq:eq14}) are poorly known, in particular for those processes that can not be experimentally studied. Therefore model calculations have to be done, in order to find these amplitudes. In case of complex amplitudes not only the absolute values are important, but also the relative signs and phases.

In this section we analyze different $t-$channel exchanges for the processes of scalar meson production in $NN$-collisions, and give, when it is possible, model independent predictions. In other words, we will find expressions for the polarization phenomena in $n+p\to n+p+S^0$, which depend only on the quantum numbers, spin and parity ${\cal J}^P$, and isospin ${\cal I}$, of the exchanged meson, but not on the corresponding coupling constants, hadronic form factors  and threshold amplitudes. Let us consider the spin structure of the matrix element (\ref{eq:eq4}) for different exchanged particles.

\subsection{Scalar exchange, ${\cal J}^P=0^+$, ${\cal I}$=0}

\begin{figure}
\mbox{\epsfysize=15.cm\leavevmode \epsffile{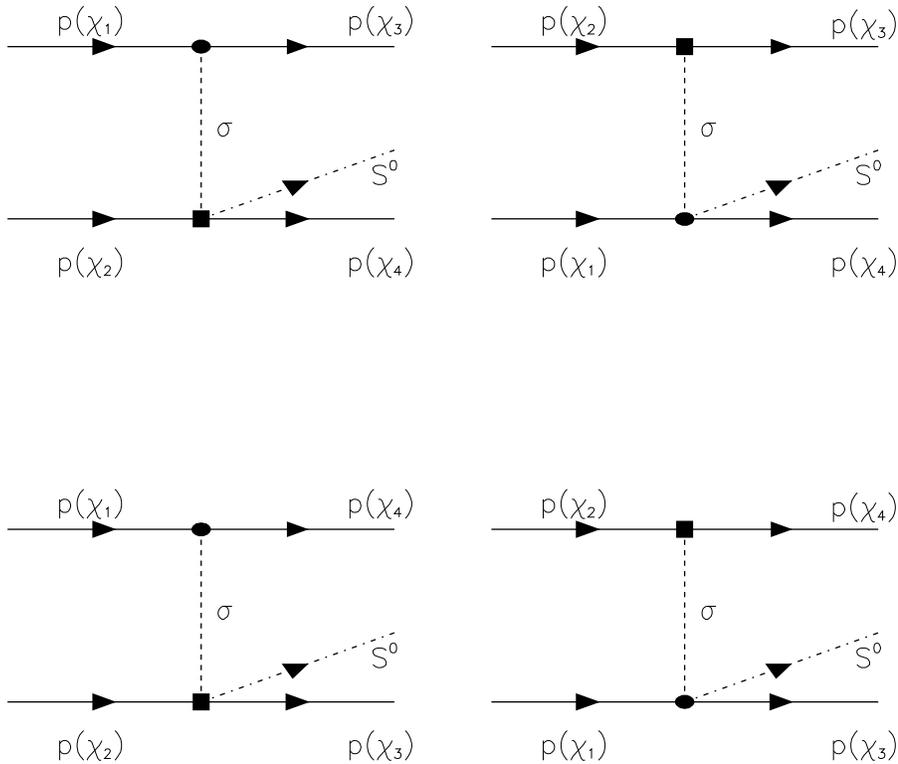}}
\vspace*{.2 truecm}
\caption{Feynman diagrams for $\sigma$-exchange, for $p+p\to p+p+S^0$, where $p(\chi_i)$, i=1-4, means that the corresponding proton (in initial or final state) is described by the 2-component spinor $\chi_i$. }
\label{fig:fig1}
\end{figure}

Taking into account the identity of initial and final protons in $p+p\to p+p+S^0$, the $\sigma$-exchange here is characterized by four different $t-$channel diagrams (Fig. \ref{fig:fig1}). Each diagram has a different spin structure, with different order of the two-component spinors $\chi_i$, $i=1-4$. Only the sum of all these diagrams, which satisfies the Pauli principle for the initial and final protons, generates the correct spin structure  (\ref{eq:mpp}), with a single amplitude $g$. In threshold conditions the spin propagator for the considered diagrams is written as:
$$\displaystyle\frac{1}{t-m^2_{\sigma}}=-\displaystyle\frac{1}
{mm_s+m^2_{\sigma}},$$
where $m$, $m_{\sigma}$, and $m_s$ are the masses of the nucleon, of the  virtual $\sigma$-meson, and of the produced scalar meson, correspondingly. Such equality is correct at the reaction threshold, and it follows from the assumption of $S$-wave production of final particles. The difference in these propagators, which appears far from threshold, generates $P$- and higher waves of  produced particles.

\begin{figure}
\mbox{\epsfysize=15.cm\leavevmode \epsffile{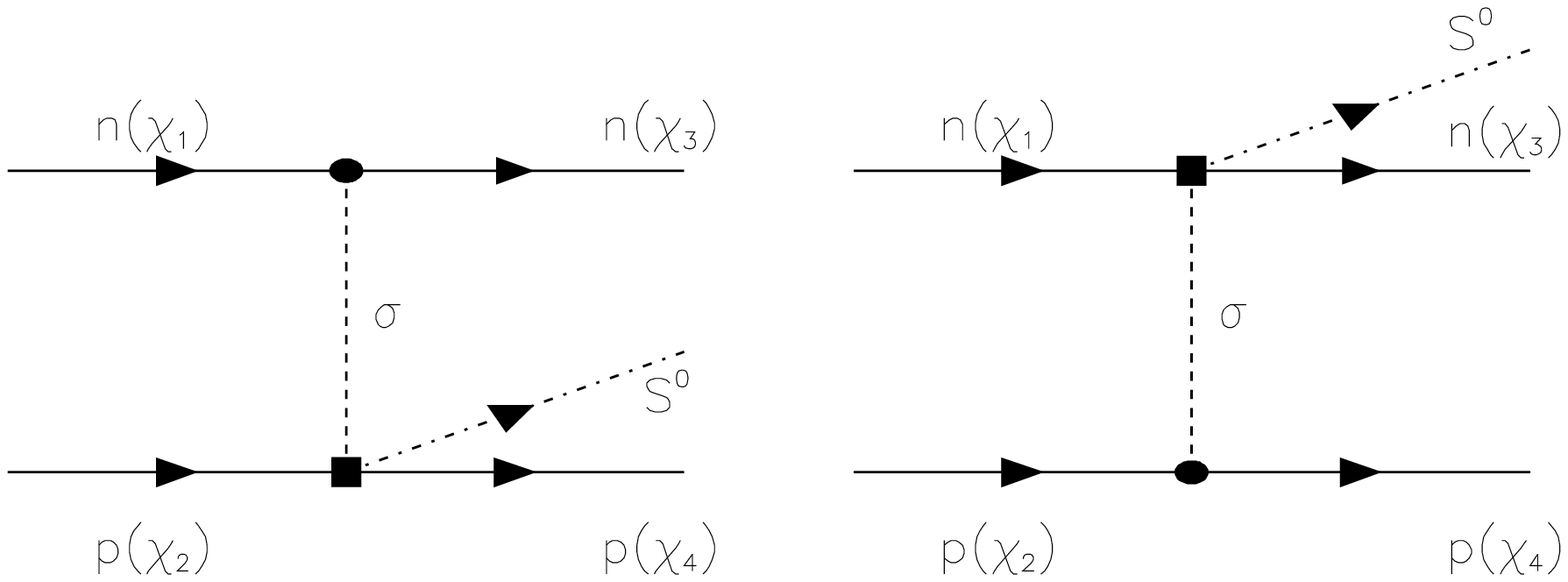}}
\vspace*{.2 truecm}
\caption{Feynman diagrams for $\sigma$-exchange for $n+p\to n+p+S^0$. Notations as in Fig. 1.}
\label{fig:fig2}
\end{figure}

The $\sigma$-exchange for  $n+p\to n+p+S^0$ is described by two diagrams, only
(Fig. \ref{fig:fig2}). Taking into account the isotopic invariance of the strong interaction, one can write the following threshold matrix element for the process $n+p\to n+p+S^0$, corresponding to $\sigma$-exchange:
$${\cal M}={\cal M}_{1\sigma}+{\cal M}_{2\sigma}=2{\cal M}_{1\sigma},$$
\begin{equation}
{\cal M}_{1\sigma}= \displaystyle\frac{g_{\sigma N N}}{t-m^2_{\sigma}}{\cal N}A(\sigma^*p\to Sp)
({\chi}^{\dagger}_4  I  {\chi}_2)({\chi}^{\dagger}_3  I {\chi}_1),
\label{eq:eq15}
\end{equation}
where $g_{\sigma N N}$ is the $\sigma N N$ coupling constant, ${\cal I} $ is the unit 2$\times$2 matrix, $ A(\sigma^*p\to Sp)$ is the partial amplitude for the subprocess 
$\sigma^*+p\to S+p$ - in the $S-$state, ${\cal N}=2m(E_1+m)=m(4m+m_s)$ is the normalization factor, related to the transformation from four-component Dirac spinors to two-component Pauli spinors.

In order to find the partial amplitudes $g_i$, corresponding to $\sigma$-exchange, it is necessary to apply the Fierz-transformation, in its two-component form, to the spinor construction in Eq. (\ref{eq:eq15}):
$$
(\chi^{\dagger}_4 I  \chi_2) ({\chi}^{\dagger}_3 I  {\chi}_1)=
-\displaystyle\frac{1}{2}(\chi^{\dagger}_4 \sigma_y\tilde\chi^{\dagger}_3)(\tilde\chi_2 \sigma_y{\chi}_1) 
+\displaystyle\frac{1}{2}(\chi^{\dagger}_4 \sigma_a\sigma_y\tilde\chi^{\dagger}_3)(\tilde\chi_2 \sigma_y\sigma_a{\chi}_1).
$$
Comparing this with the general parametrization of the spin structure for the threshold matrix element of the process 
$n+p\to n+p+S^0$, one can find the following expressions for the particle amplitudes $g_i$:
\begin{eqnarray}
&g_{1\sigma}(np\to npS^0)=&-A_{\sigma}, \nonumber  \\
&g_{2\sigma}(np\to npS^0)=&A_{\sigma},\label{eq:eq16} \\
&g_{3\sigma}(np\to npS^0)=&A_{\sigma}, \nonumber  \\
&g_{\sigma}(pp\to ppS^0)=&-2A_{\sigma}, \nonumber  
\end{eqnarray}
where
$$A_{\sigma}={\cal N}A(\sigma p\to pS^0)\displaystyle\frac{g_{\sigma N N}}{t-m^2_{\sigma}}.$$

Eqs. (\ref{eq:eq16}) allow to predict all polarization phenomena in $n+p\to n+p+S^0$, independently on the details of the considered model, such as the value of the constant $g_{\sigma N N}$ and the amplitude $A(\sigma p\to pS^0)$.

One can find:
\begin{equation}
p_{1\sigma}=1,~A_{1\sigma}=A_{2\sigma}=p_{2\sigma}=p_{3\sigma}=p_{4\sigma}=0,
\label{eq:eq17}
\end{equation}
\begin{equation}
{\cal R}_{\sigma}=\displaystyle\frac{\sigma(np\to npS^0)}{\sigma(pp\to ppS^0)}=2.
\label{eq:eq18}
\end{equation}
In (\ref{eq:eq18}) we took into account the identity of the produced protons in the reaction $p+p\to p+p+S^0$.

\subsection{$\eta$-exchange, ${\cal J}^P=0^-$, ${\cal I}$=0}

As in the case of $\sigma$-exchange, the $\eta$-exchange generates four different diagrams for $pp$-interaction and two diagrams for $np$-interaction. Let us consider, therefore, the simplest case of $n+p\to n+p+S^0$-processes, the matrix element of which can be written as:
$$
{\cal M}={\cal M}_{1\eta}+{\cal M}_{2\eta}=2{\cal M}_{1\eta},$$
\begin{equation}
{~\cal M}_{1\eta}=-A_{\eta}({\chi}^{\dagger}_4\vec\sigma\cdot\hat{\vec k }\chi_2)(\chi^{\dagger}_3\vec\sigma\cdot\hat{\vec k }\chi_1,)
\label{eq:eq19}
\end{equation}
$$A_{\eta}=\displaystyle\frac{{\cal N} k}{E+m}~\displaystyle\frac{g_{\eta N N}}{t-m^2_{\eta}}~A(\eta^*N\to NS^0),$$
where $g_{\eta N N}$ is the coupling constant for the $\eta NN$-vertex, $A(\eta^*N\to N S^0)$ is the threshold amplitude for $\eta+N\to N+S^0$, which describes the partial transition: $\ell_{\eta }=1\to j^P=1/2^+$. The kinematical factor $\displaystyle\frac{k}{E+m}=\sqrt{\displaystyle\frac{m_s}{4m+m_s}}$ (at threshold) results from the transformation of the pseudoscalar vertex $\overline{u}\gamma_5u$ to its two-component equivalent $\chi^{\dagger}\vec\sigma\cdot \hat{\vec k }\chi$.

After applying the Fierz transformation to the spin structure (\ref{eq:eq19}) to the standard form,  
Eq. (\ref{eq:eq4}), one can find the following formulas for the partial amplitudes $g_{i\eta}$, $i=1-3$, corresponding to $\eta$-exchange:
\begin{eqnarray}
&g_{1\eta}(np\to npS^0)=&-A_{\eta}, \nonumber  \\
&g_{2\eta}(np\to npS^0)=&-A_{\eta},\label{eq:eq20} \\
&g_{3\eta}(np\to npS^0)=&A_{\eta}, \nonumber  \\
&g_{\eta}(pp\to ppS^0)=&-2A_{\eta}, \nonumber  
\end{eqnarray}
and similarly, to $\sigma$-exchange, one can predict all polarization observables:
\begin{equation}
A_{1\eta}=A_{2\eta}=p_{3\eta}=p_{4\eta}=0,~p_{1\eta}=-1,~p_{2\eta}=2, {\cal R}=2.
\label{eq:eq21}
\end{equation}

Comparing Eq. (\ref{eq:eq17}) and Eq.(\ref{eq:eq21}), one can see that the polarization transfer coefficients, characterizing a change of neutron polarization, must be very sensitive to the quantum number of the isoscalar exchange in $t-$channel, other polarization observables, such as ${\cal A}_1$, ${\cal A}_2$, $p_3$ and $p_4$ vanish for both exchanges ${\cal J}^P=0^+$ and ${\cal J}^P=0^-$.

\subsection{Pion exchange, ${\cal J}^P=0^-$, ${\cal I}$=1}

\begin{figure}
\mbox{\epsfysize=15.cm\leavevmode \epsffile{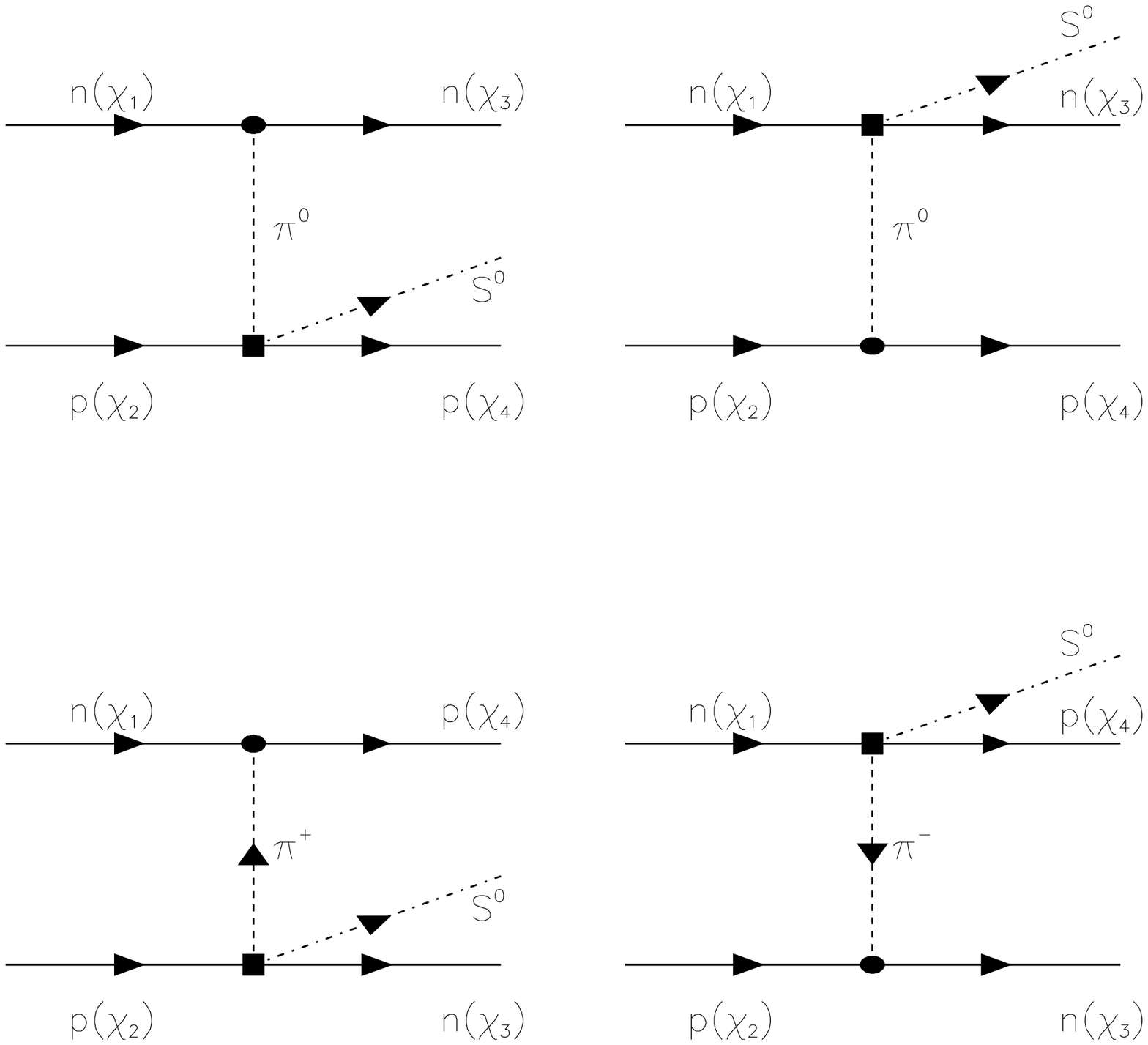}}
\vspace*{.2 truecm}
\caption{Feynman diagrams for $\pi$-exchange for $n+p\to n+p+S^0$. Notations as in Fig. 1.}
\label{fig:fig3}
\end{figure}

Any isovector exchange for the process $n+p\to n+p+S^0$ is characterized by a set of four Feynman diagrams Fig. (\ref{fig:fig3}) with exchange of neutral and charged pion.

Considering the contribution of all these diagrams, and taking into account the isotopic invariance predictions for the two vertices of the considered diagrams:
$$g_{\pi^0 pp}=-g_{\pi^0 nn}=
\displaystyle\frac{1}{\sqrt{2}}g_{\pi^+ np}=
\displaystyle\frac{1}{\sqrt{2}}g_{\pi^- pn},$$
one can find (after applying the Fierz transformation) the following expressions for the partial amplitudes 
$g_{i\pi}$, $i=1-3$, corresponding to $\pi$-exchange:
\begin{eqnarray}
&g_{1\pi}(np\to npS^0)=&A_{\pi}, \nonumber  \\
&g_{2\pi}(np\to npS^0)=&-3A_{\pi},\label{eq:eq22} \\
&g_{3\pi}(np\to npS^0)=&3A_{\pi}, \nonumber  \\
&g_{\pi}(pp\to ppS^0)=&2A_{\pi}, \nonumber  
\end{eqnarray}
where $A_{\pi}={\cal N}\sqrt{\displaystyle\frac{m_s}{4m+m_s}}\displaystyle\frac{g_{\pi N N}}{t-m^2_{\pi}}A(\pi^*N\to NS).$ In principle, the amplitude $A(\pi^*N\to NS)$ for threshold $S$-production in $\pi N$-interaction can be experimentally measured through the corresponding differential cross section, but this method will not determine its sign.

However numerical values of the polarization phenomena for the $n+p\to n+p+S^0$ processes can be predicted exactly, without information on $A(\pi^*N\to NS)$:
\begin{equation}
A_{1\pi}=2/7,~A_{2\pi}=0,~p_{1\pi}=-3/7,~p_{2\pi}=6/7,~p_{3\pi}=-6/7,~p_{4\pi}=12/7,
\label{eq:eq23}
\end{equation}
and they are very different from the case of isoscalar exchange.

Another interesting result concerns the large difference in cross section, for $pp$- and $np$-interaction: ${\cal R}_{\pi}=14$, i.e. for scalar meson production (in $NN-$interaction) we have large isotopic dependence.

\subsection{Scalar exchange, ${\cal J}^P=0^+$, ${\cal I}$=1}

In this case, both processes $p+p\to p+p+S^0$ and $n+p\to n+p+S^0$ is characterized by a set of four Feynman diagrams, the neutral $a_0$-exchange for $pp-$collisions and the neutral+charged $a_0$-exchanges for $np$-collisions.

The partial amplitudes 
$g_{ia}$, $i=1-3$, corresponding to $a_0$-exchange are determined by the following formulas:
\begin{eqnarray}
&g_{1a}(np\to npS^0)=&A_{a}, \nonumber  \\
&g_{2a}(np\to npS^0)=&3A_{a},\label{eq:eq24} \\
&g_{3a}(np\to npS^0)=&3A_{a}, \nonumber  \\
&g_{a}(pp\to ppS^0)=&2A_{a}, \nonumber  
\end{eqnarray}
and $A_{a}={\cal N}\displaystyle\frac{g_{a N N}}{t-m^2_a}A(a^*N\to NS)$. The threshold amplitude $A(a^*N\to NS)$ for the exotic subprocess $a^*+N\to N+S^0$ has to be determined in framework of a model. 

From these formulas, independently from the values of the coupling constant $g_{a N N}$ and the threshold amplitude $A(a^*N\to NS)$, one can predict the following values for the polarization observables:
\begin{equation}
A_{1a}=2/7,~A_{2a}=0,~p_{1a}=3/7,~p_{2a}=0,~p_{3a}=6/7,~p_{4a}=0,~{\cal R}_{a}=14.
\label{eq:eq25}
\end{equation}
They are  different from the case of isoscalar exchange and from (\ref{eq:eq25}). Due to isovector exchange, the isotopic ratio ${\cal R}$ is large also in this case.

\subsection{Correlation of spin and isospin structures for the threshold regime}

Combining the previously derived contributions, one can find for the threshold amplitudes $g_i$ the following expressions:
\begin{eqnarray}
&g_{1}(np\to npS^0)=&-A_{\sigma}+A_{a}-A_{\eta}+A_{\pi}, \nonumber  \\
&g_{2}(np\to npS^0)=& A_{\sigma}+3A_{a}-A_{\eta}-3A_{\pi},\label{eq:eq26} \\
&g_{3}(np\to npS^0)=& A_{\sigma}+3A_{a}+A_{\eta}+3A_{\pi}, \nonumber  \\
&g(pp\to ppS^0)=& 2g_1(np\to npS^0). \nonumber  
\end{eqnarray}

Let us note that the S-channel parametrization of the spin structure for the threshold matrix elements allows (after applying the Fierz transformation) to unify different $t-$channel contributions- with different ${\cal J}^P$ - in universal and transparent form, which is well adapted to the analysis of the sensitivity of the  polarization phenomena, in $n+p\to n+p+S^0$, to the quantum numbers of the $t-$channel meson. Such unification allows to simplify all calculations: we can express the matrix element for any exchange in terms of three amplitudes only, whereas, for example, a model with $\eta+\pi+\sigma+a$ contains $2+4+2+4=12$ different Feynman diagrams - with different spin structures. All these twelve contributions to the total matrix elements, in general, interfere, so it is in principle necessary to calculate $12\times 12=144$ terms, instead than $3\times 3=9$ terms, for the $g_i$ parametrization of the matrix elements. Moreover, these nine terms are the same for any model, whereas, for example, adding vector exchanges will increase essentially the number of $t-$contributions. Another advantage of the $t\to s$ Fierz transformation is the explicit dependence of the observables on a definite combination of coupling constants, hadronic form factors and elementary amplitudes. This helps in finding out which contributions play the most important role in the $t-$channel dynamics, and gives a feedback on the coupling constants by comparison with the experimental data.

Another important property of the partial amplitudes $g_i$ for $np-$processes, Eq. (\ref{eq:eq26}), concerns a strong correlation of the spin and isospin structure of the matrix element in the threshold region. The amplitudes $g_2(np\to npS^0)$ and $g_{3}(np\to npS^0)$, describing the $np$-interaction in the isotopic singlet state, are in general different, even in case of a definite isospin in t-channel, if the P-parity of the $t$-channel is not fixed. However, for definite ${\cal J}^P$-exchanges, we have:
\begin{eqnarray}
&g_{2}(np\to npS^0)=&g_{3}(np\to npS^0),~\mbox{if~} {\cal J}^P=0^+,~{\cal I}=0 ~\mbox{and~}{\cal I}=1,\nonumber  \\
&g_{2}(np\to npS^0)=&-g_{3}(np\to npS^0),~\mbox{if~} {\cal J}^P=0^-,~{\cal I}=0 ~\mbox{and~}{\cal I}=1.
\label{eq:eq27} 
\end{eqnarray}
The relations (\ref{eq:eq27}) hold also for any combination of isoscalar and isovector exchanges.

The total cross section for $n+p\to n+p+S^0$ is the incoherent sum of isosinglet and isotriplet contributions:
\begin{eqnarray}
&\sigma(np\to npS^0)=&\sigma_0(np\to npS^0)+\sigma_1(np\to npS^0),\nonumber \\
&\sigma_0(np\to npS^0)=&2|g_2(np\to npS^0)|^2+|g_{3}(np\to npS^0)|^2,\label{eq:eq28}\\
&\sigma_1(np\to npS^0)=&|g_1(np\to npS^0)|^2= \displaystyle\frac{1}{2}\sigma(pp\to ppS^0).\nonumber 
\end{eqnarray}

\section{A possible model for $N+N\to N+N+S^0$ }
We analyze here in detail a model for $N+N\to N+N+S^0$, based on $\sigma+\pi$-exchange. In order to justify such model, let us mention that the $\eta$-contribution can be neglected, in (\ref{eq:eq26}), due to the fact that the coupling constant $g_{\eta N N}$ is presently poorly known, being in the range 1$\div 7$ \cite{Be99,Ma89,Be95,Ch02}. Concerning the $a_0$-exchange, both the ingredients of such contribution, the coupling constant $g_{a N N}$ and the amplitude $A(aN\to NS^0)$ are presently not known. The mass of the $a_0$-meson is larger in comparison with the pion mass.  In any case, the $\sigma+\pi$-model can be considered a realistic starting point for the analysis, containing the exchange of mesons with different P-parities and isospin. As a result, all three partial amplitudes are different:
\begin{eqnarray}
&g_{1}(np\to npS^0)=&A_{\pi}(1-r), \nonumber  \\
&g_{2}(np\to npS^0)=&A_{\pi}(-3+r),\label{eq:eq29} \\
&g_{3}(np\to npS^0)=&A_{\pi}(3+r), \nonumber  \\
&g(pp\to ppS^0)=&2A_{\pi}(1-r), \nonumber  
\end{eqnarray}
where $r$ is the ratio of the corresponding contributions:
$$r=\displaystyle\frac{g_{\sigma N N}}{g_{\pi N N}}\left ( \displaystyle\frac{t-m^2_{\sigma}}{t-m^2_{\pi}}\right )
\displaystyle\frac{A(\sigma^*N\to NS^0)}{A(\pi^*N\to NS^0)}
\sqrt{\left (1+4\displaystyle\frac{m}{m_s}\right )}.$$
One can see that all the physics of this model (with eight different Feynman diagrams) is contained in a single parameter $r$, which is basically the ratio of the coupling constants and the elementary amplitudes. In the general case, the ratio $r$ is a complex parameter, which depends on the excitation energy of the produced $NNS$-system. Therefore the polarization phenomena for the process $n+p\to n+p+S^0$ and the ratio of cross section for $np$- and $pp$-collisions, can be expressed in terms of two parameters $|r|^2$ and 
${\cal R}e r$:
\begin{equation}
{\cal R}=\displaystyle\frac{\sigma(np\to npS^0)}{\sigma(pp\to ppS^0)}=2\displaystyle\frac{7-2{\cal R}e r+|r|^2}{1-2{\cal R}e r+|r|^2}=2\left ( 7-6 \displaystyle\frac{|r|^2-2{\cal R}e r}{1-2{\cal R}e r+|r|^2}
\right ).
\label{eq:eq30}
\end{equation}
The coefficients ${\cal A}_1$ and ${\cal A}_2$, which characterize the polarized $\vec n\vec p$-collisions, can be written as follows:
\begin{equation}
{\cal A}_1=2\displaystyle\frac{1+{\cal R}e r}{ 7-2{\cal R}e r+|r|^2},~{\cal A}_2=
\displaystyle\frac{3{\cal R}e r}{ 7-2{\cal R}e r+|r|^2}.
\label{eq:eq31}
\end{equation}

The ratio $r$ can be found in framework of a model for the elementary subprocesses $\pi^*(\sigma^*)+N\to N+S^0$, under several assumptions concerning coupling constants, cut-off parameters, form factors... To avoid the uncertainties related to these choices, we assume that $r$ is real. This is the case, for example, of the effective Lagrangian approach, which gives real amplitudes and, therefore, real values for the ratio $r$. Another possibility is the saturation of the $A(\sigma^*N\to NS^0)$ and $A(\pi^*N\to NS^0)$ amplitudes by a single nucleon resonance with ${\cal J}^P=1/2^+$. Such mechanism produce complex amplitudes with zero relative phase, and $r$ is a real parameter, too.

In this case we can predict the $r$-dependence of ${\cal R}$, ${\cal A}_1$, and ${\cal A}_2$ (Figs. 4-6). The ratio $R$ of the corresponding cross sections can be very large in the region $r\simeq 1$ (Fig. 4). The experimental determination of $R$ will allow to find two solutions for $r$:
$$r_{\pm}=1\pm 2\sqrt{\displaystyle\frac{3}{R-2}}$$

The function $A_1(r)$ has two extrema: the maximum at $r=-1+\sqrt{10}$, where $A_1^{max}=(2+\sqrt{10})/\sqrt{6}$
and a minimum at $r=-1-\sqrt{10}$, where $A_1^{min}=(2-\sqrt{10})/\sqrt{6}$, see Fig. 5.

The function $A_2(r)$ has a very similar behavior, with a  the maximum at $r=+\sqrt{7}$, where $A_2^{max}=(1+\sqrt{7})/4$
and a minimum at $r=-\sqrt{7}$, where $A_2^{min}=12-\sqrt{7})/4$, see Fig. 6.

\section{Comments on the NN final state interaction (FSI)}

In order to illustrate possible effects of FSI for the considered reaction, let us make an oversimplified estimation, using only the $NN$-scattering length approximation, which will result in an upper limit for the NN-FSI. Let us consider, for illustration, the case of pion exchange. One finds the following correction for some observables which have been previously discussed:

\begin{itemize}
\item The ratio of $np$ and $pp$-cross sections:
$$R^{(\pi)}=\displaystyle\frac{1}{2}\left [ \displaystyle\frac{a_s(np)}{a_s(pp)}\right ]^2+
\displaystyle\frac{27}{2}\left [ \displaystyle\frac{a_t(np)}{a_s(pp)}\right ]^2.$$
\item Polarization effects for $np$-collisions:
$$C_{nn}^{(\pi)}=\displaystyle\frac{9-R_{st}^2}{27+R_{st}^2},~
D_{nn}^{(\pi)}=\displaystyle\frac{6(R_{st}-3)}{27+R_{st}^2},$$
where $R_{st}=[a_s(np)/a_t(np)]$, $a_s(NN)$ is the singlet $NN$-scattering length and
$a_t(np)$ is the triplet $np$-scattering length (for S-state).
\end{itemize}

Using the following values for the scattering lengths \cite{Du83}: $a_s(np)$=-23.768 fm, 
$a_t(np)$=5.424 fm, and $a_s(pp)$ =-7.8098 fm, one can find:
$R^{(\pi)}\simeq 11$ (instead of 14), $C_{nn}^{(\pi)}\simeq $= -0.22 (instead of 0.29), and $D_{nn}^{(\pi)}\simeq $-0.96 (instead of -0.43), 
i.e. the final state $np$-interaction can change the sign of $C_{nn}^{(\pi)}$, but does not affect much the ratio $R^{(\pi)}$.

\section{Conclusions}
Let us summarize the main results concerning the production of scalar mesons in nucleon-nucleon collisions near threshold.

\begin{itemize}
\item We parametrize the spin structure of the threshold matrix element  for the processes $p+p\to p+p+S^0$, where $S^0$ is a neutral scalar meson, $S= \sigma$, $f_0$, or $a_0$, in terms of a single spin structure, corresponding to a singlet-singlet transition in the $pp-$system. The same structure describes the production of the charged $a_0^{\pm}$-mesons in the processes: $p+p\to p+n+a_0^+$ and $n+p\to p+p+a_0^-$.

\item The process $n+p\to p+p+S^0$ is characterized by a more complicated spin structure, where the matrix element contains, in the general case, three contributions with different spin structures. One contribution coincides with the $ {\cal M}(pp\to pp S^0)$ matrix element -due to isotopic invariance, and two additional contributions describe the triplet-triplet transition in the $np$-system, which are forbidden by the Pauli principle, in case of the $pp-$system.

\item The suggested model-independent parametrization of the spin structure, which is based on the general symmetry properties of the strong interaction ( such as P-invariance, isotopic invariance, conservation of angular momentum and Pauli principle) allows to analyze polarization effects for $pp$- and $np-$collisions in a transparent way. These effects are very peculiar for threshold conditions, where all final particles are emitted in relative $S$-states.

\item The standard $t-$channel dynamics for the process $N+N\to N+N+S^0$ is generated by different meson exchanges with ${\cal J}^{ P}=0^+$ and 
${\cal J}^P=0^-$ and isotopic spin ${\cal I}$=0 and ${\cal I}$=1.
The dependence of the polarization phenomena in $n+p\to n+p+S^0$ on the quantum number of the exchanged mesons can be described in terms of the corresponding S-channel partial amplitudes, with the help of the Fierz transformation in its two-component form.
\item As an example of a 'realistic' model in the threshold region, we considered the $\pi+\sigma$-exchange. An attractive property of such model is that the polarization phenomena in $np$-collisions depend on a single complex parameter $r$, which characterizes the relative role of $\sigma$- and $\pi$-exchange. The polarization observables, which do not vanish, show a large sensitivity to $r$.
\item The processes $p+p\to p+p+S^0$ and $n+p\to n+p+S^0$ show large isotopic effects, i.e. S-state production of scalar mesons is larger in $np$-collisions  than in $pp$-collisions, and polarization phenomena are different for these processes.
\end{itemize}

Note that these results do not depend on the structure of the $f_0$ and $a_0$-mesons ($q\overline{q}$-states, four-quark state, $K\overline{K}$-molecule etc.) and on the mechanisms of their decays.

\begin{figure}
\mbox{\epsfysize=15.cm\leavevmode \epsffile{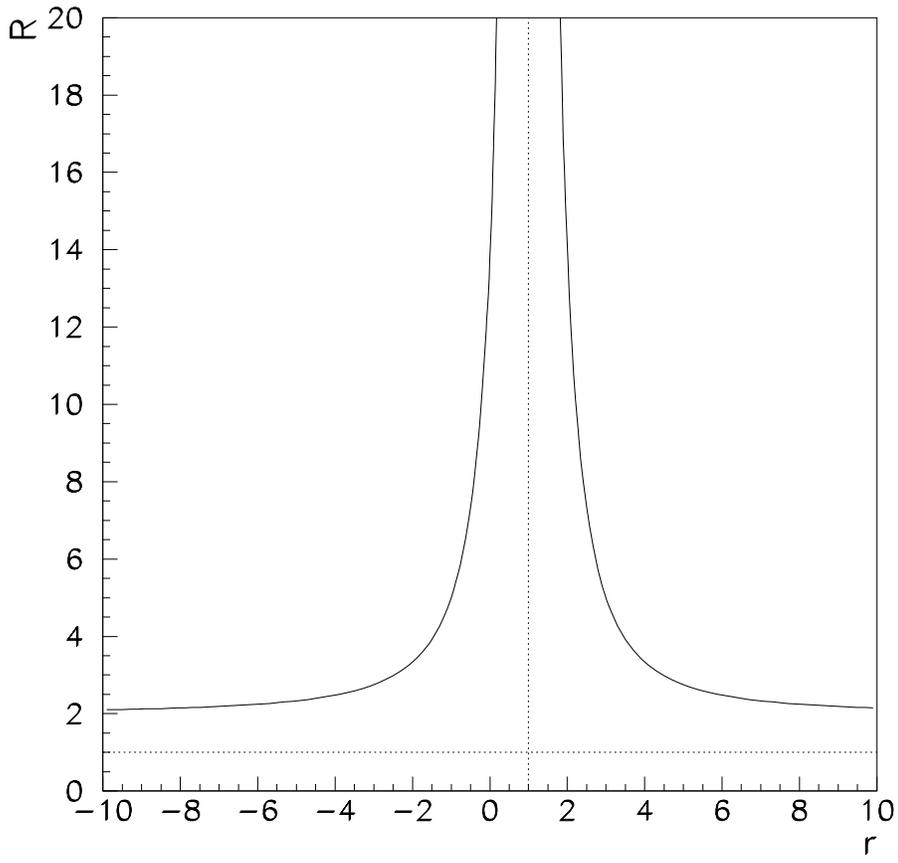}}
\vspace*{.2 truecm}
\caption{Dependence of $R$ on $r$, see Eq. 30}
\label{fig:fig4}
\end{figure}

\begin{figure}
\mbox{\epsfysize=15.cm\leavevmode \epsffile{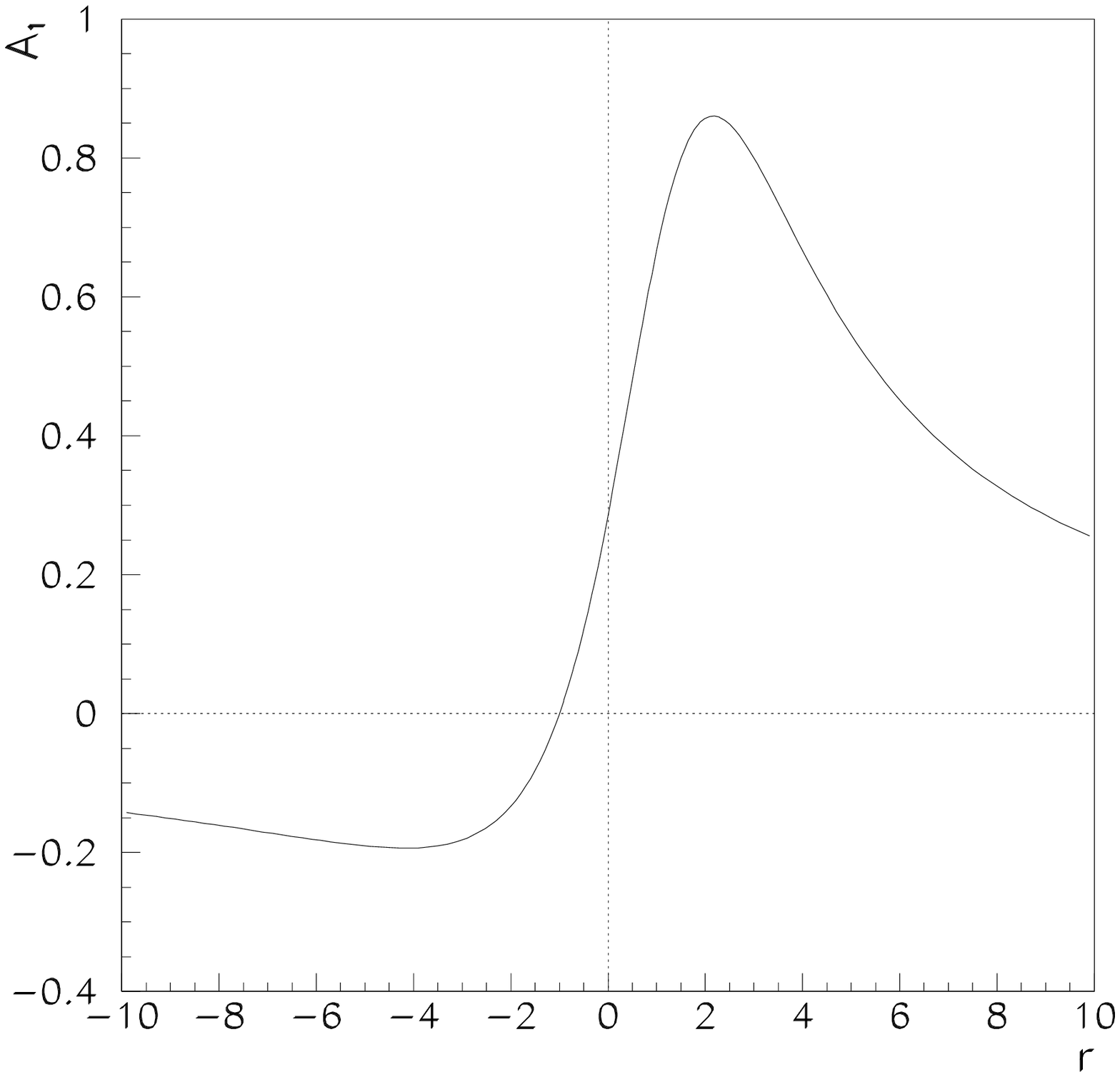}}
\vspace*{.2 truecm}
\caption{Dependence of ${\cal A}_1$ on $r$, see Eq. (31)}
\label{fig:fig5}
\end{figure}
\begin{figure}
\mbox{\epsfysize=15.cm\leavevmode \epsffile{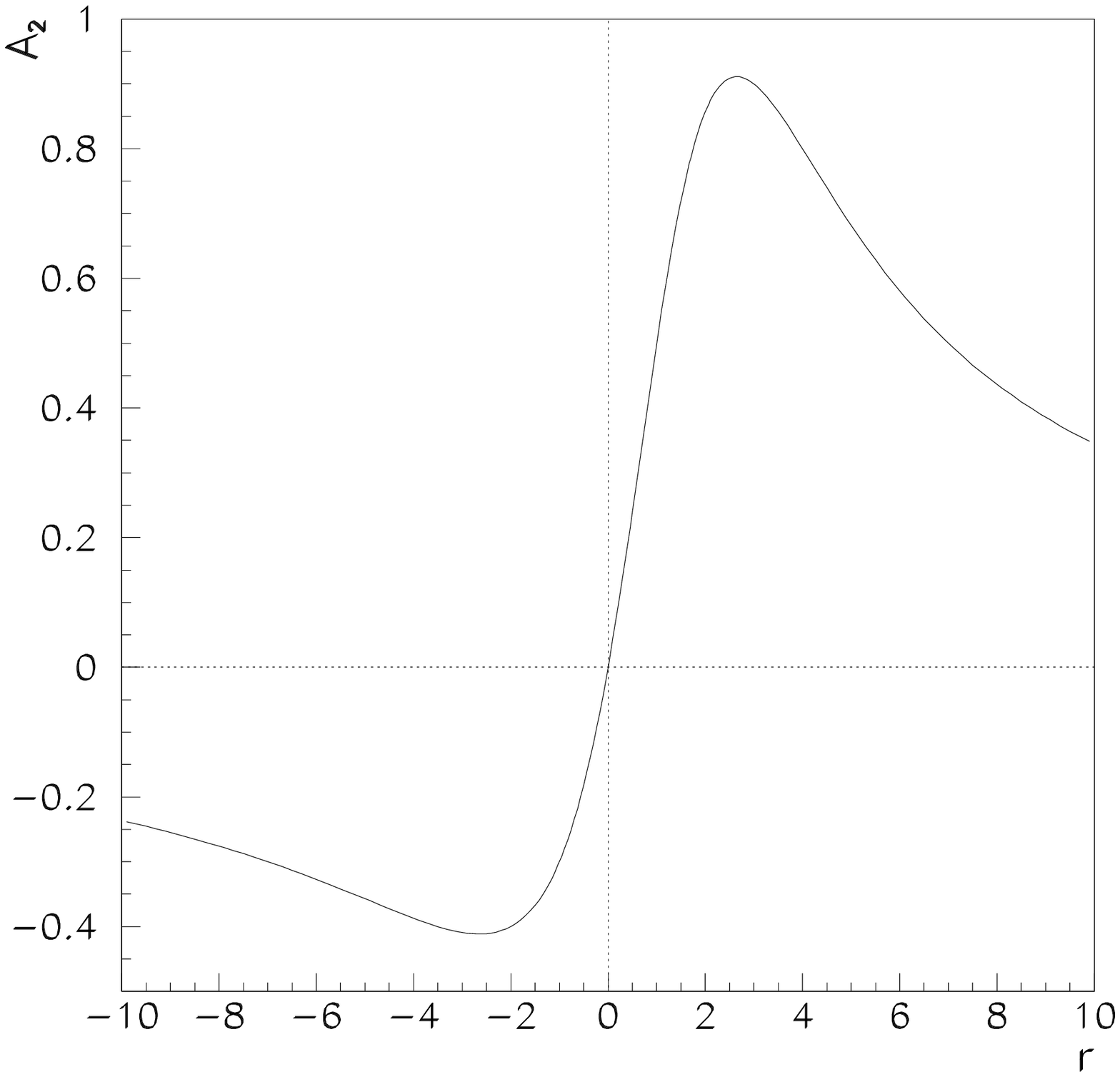}}
\vspace*{.2 truecm}
\caption{Dependence of ${\cal A}_2$ on $r$, see Eq. (31).}
\label{fig:fig6}
\end{figure}
\end{document}